\documentclass{elsart}
\usepackage[dvips]{epsfig}

\journal{Nuclear Instruments and Methods A}

\begin{document}
{\tt DESY 00-065} \hfill {\tt physics/0004063}\\
{\tt April 2000}  \hfill
\vfill
\small
\begin{frontmatter}
\title {Test Results on the Silicon Pixel Detector for the 
        TTF-FEL Beam Trajectory Monitor}

\author[uni]{S.~Hillert},
\author[desy]{R.~Ischebeck}, 
\author[desy]{U.~C.~M\"uller},
\author[desy]{S.~Roth},
\author[desy]{K.~Hansen},
\author[ketek]{P.~Holl},
\author[desy]{S.~Karstensen}, 
\author[ketek]{J.~Kemmer},
\author[uni,desy]{R.~Klanner},
\author[ketek]{P.~Lechner},
\author[desy]{M.~Leenen},
\author[desy]{J.~S.~T.~Ng\thanksref{slac}},
\author[uni]{P.~Schm\"user},
\author[mpe]{L.~Str\"uder}

\address [uni]  {II.\ Institut f\"ur Experimentalphysik, 
                 Universit\"at Hamburg, Luruper Chausee 149,\\
                 D-22761 Hamburg, Germany} 
\address [desy] {Deutsches Elektronen-Synchrotron DESY, 
                 Notkestra{\ss}e 85,\\
                 D-22603 Hamburg, Germany} 
\address [ketek]{Ketek GmbH, Am Isarbach 30,\\
                 D-85764 Oberschlei{\ss}heim, Germany}
\address [mpe]  {Max-Planck-Institut f\"ur extraterrestrische Physik,
                 Giessenbachstra{\ss}e,\\
                 D-85740 Garching, Germany}


\thanks  [slac] {now at SLAC, Stanford, CA 94309, USA}

\begin{abstract}
Test measurements on the silicon pixel detector for the beam trajectory
monitor at the free electron laser of the TESLA test facility are presented.
To determine the electronic noise of detector and read-out
and to calibrate the signal amplitude of different pixels
the 6 keV photons of the manganese $K_{\alpha}/K_{\beta}$ line are used.
Two different methods determine the spatial accuracy of the detector:
In one setup a laser beam is focused to a straight line and moved across 
the pixel structure.
In the other the detector is scanned using a low-intensity electron beam of an 
electron microscope.
Both methods show that the symmetry axis of the detector defines a straight 
line within 0.4 $\mu$m.
The sensitivity of the detector to low energy X-rays is measured 
using a vacuum ultraviolet beam at the synchrotron light source HASYLAB. 
Additionally, the electron microscope is used to study the radiation hardness 
of the detector.
\end{abstract}

\begin{keyword}
Beam monitor,
X-ray detector, 
solid-state detector, 
imaging sensor.
\PACS{41.85.Qg, 07.85.Fv, 29.40.Wk, 42.79.Pw.}
\end{keyword}

\end{frontmatter}

\section{Introduction}

It is a widespread opinion that the fourth generation synchrotron
light source will be a X-ray free electron laser (FEL).
It will consist of a single-pass FEL relying on the
self-amplified spontaneous emission (SASE) mechanism \cite{sase}
and deliver coherent radiation in the X-ray range with unprecedented 
brilliance.
In such SASE FEL a tightly collimated electron beam of high charge
density is sent through a long undulator.
The SASE effect results from the interaction of the electron bunch
with its own radiation field created by the undulation motion.
This interaction can only take place if the electron and the 
photon beams overlap.

\begin{figure*}
 \begin{center}
 \mbox{\epsfig{figure=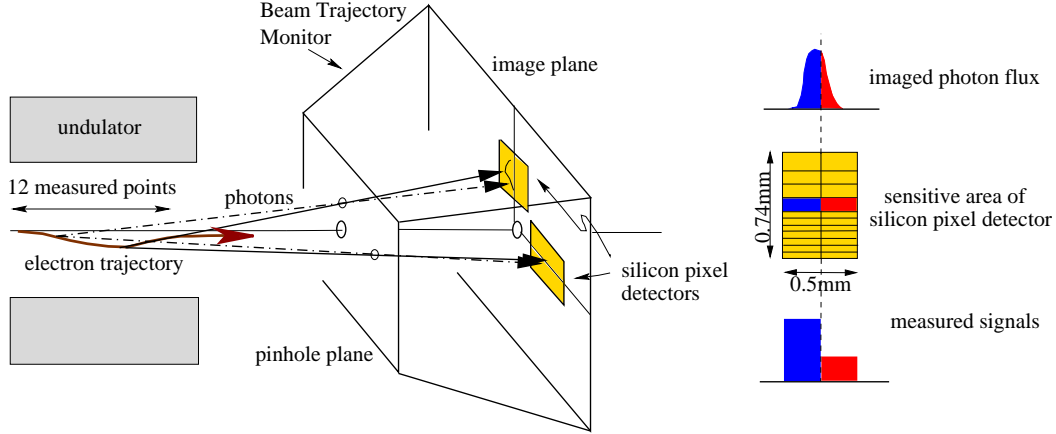,%
               width=\textwidth}}
 \end{center}
 \caption{Measurement principle: 
          An image of the electron beam is projected through a set of
          pinholes onto pixel detectors.}
 \label{fig_principle}
\end{figure*}

At the free electron laser of the TESLA Test Facility (TTF-FEL) 
\cite{ttf_fel_prop} the electron beam position must be controlled 
to better than 10~$\mu$m over the entire 15~m long undulator.
With the beam trajectory monitor (BTM) \cite{btm_prop}
the off-axis spontaneous undulator radiation is imaged through a set of
pinholes of 80~$\mu$m diameter (see Fig.~\ref{fig_principle}).
In order to reduce the effect of diffraction, only the higher
harmonics of the spontaneous undulator radiation will be used 
for BTM measurements.
A 120~nm thick silver foil across each pinhole absorbs all 
low energy photons and restricts the spectral range of the 
detected radiation to energies above 100~eV.
From a simulation using the expected undulator spectrum the Gaussian 
width of the photon spot at the position of the detector
0.5~m behind the pinholes is estimated to 30~$\mu$m.
To achieve the required resolution of the BTM, the center of the
photon spot will be measured with a precision of better than 1~$\mu$m
using a high resolution silicon pixel detector.
It delivers 12 points of the transverse beam position with an
accuracy of better than 10~$\mu$m over a length of 5~m using a single setup.
The performance of the silicon detector with respect to noise, 
spatial precision, radiation hardness and quantum efficiency is presented 
in this paper.

\section{The Silicon Pixel Detector}

\begin{figure}[t]
 \begin{center}
 \mbox{\epsfig{figure=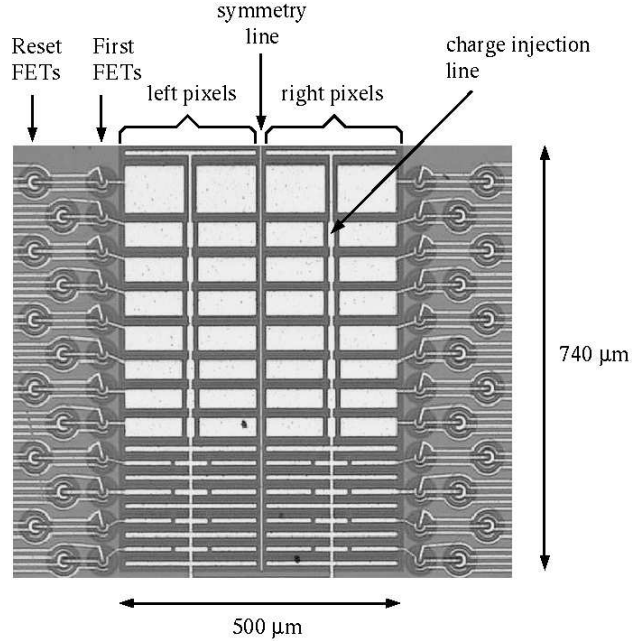,%
               width=0.6\textwidth,clip=}}
 \end{center}
 \caption{Anode structure of the silicon pixel detector:
          Two pixel rows with a charge injection line across each row
          and each pixel connected to the on-chip JFETs}
 \label{fig_lobster}
\end{figure}

A silicon pixel detector with an active area of 0.5 mm $\times$ 0.74 mm 
and a total of 24 channels was designed and fabricated at the 
MPI Halbleiterlabor.
The sensitive area of the detector consists of two rows of each 12 active 
pixels as shown in Fig.~\ref{fig_lobster}.  
Each pixel anode is directly connected to an on-chip JFET for 
low noise read-out.
The pixels are 250~$\mu$m wide, with heights varying from 25~$\mu$m 
(nearest to beam) to 100~$\mu$m to give roughly equidistant measuring 
points in the projection along the undulator axis.
High quantum efficiency is achieved using a thin entrance window technology
\cite{hll_ketek}.

\begin{figure*}[t]
 \begin{center}
 \mbox{\epsfig{figure=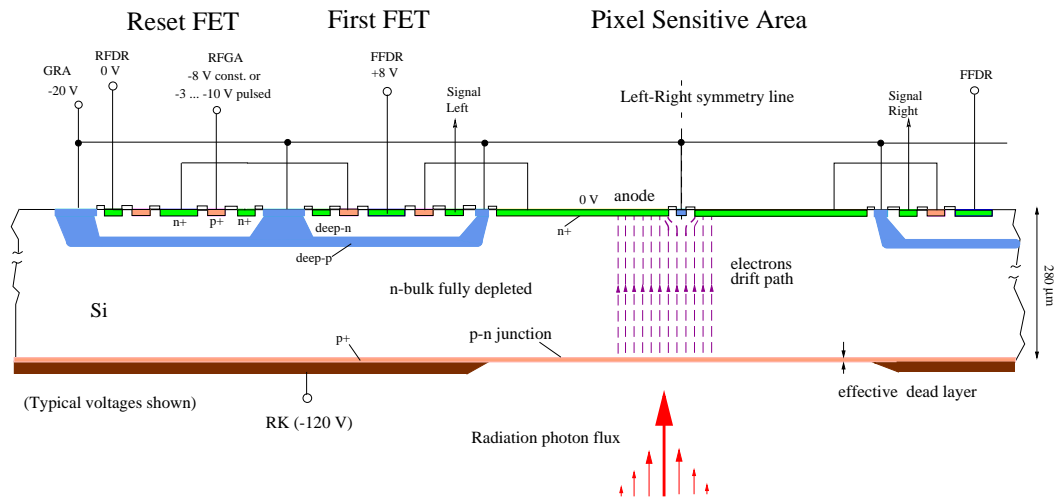,%
               width=\textwidth}}
 \end{center}
 \caption{Cross section of the pixel detector.}
 \label{fig_lobxsec}
\end{figure*}

The concept of a backside illuminated $pn$-junction detector has been 
chosen, which shows not only a high quantum efficiency for the desired
photon energies, but in addition an excellent spatial homogeneity.
It consists of a fully depleted $n$-type bulk and a non structured
$p^+$-rear contact, acting as radiation entrance window.
At photon energies of about 150~eV the absorption length in silicon drops 
to 40~nm, which leads to signal loss in the almost field free, highly
doped region underneath the $p^+$ contact.
To reduce the width of this insensitive region the implantation
of boron ions was done through a SiO$_2$ layer, which has been removed 
afterwards.
One achieves a shallow doping profile with the $pn$-junction placed
at a depth of only 30~nm below the detector surface.
Ionizing radiation which penetrates through the dead layer generates
electron hole pairs in the sensitive volume of the detector.
Applying a negative voltage of about 120~V to the rear contact totally 
depletes the detector and causes the signal electrons
to drift towards the collecting $n^+$ pixel anodes
(see Fig.~\ref{fig_lobxsec}).

The pixels are formed by $n^+$-implants and are isolated from each other
by a 5~$\mu$m wide $p^+$ grid.
Each pixel anode is connected to an amplifying JFET which is integrated
on the detector chip, thus minimizing stray capacitances.
The JFETs are operated as source followers with a given constant current
of about 200~$\mu{}$A from source to drain.
The collected signal charge is transferred to the gate and modulates
the voltage drop across the JFET.
A second JFET (Reset FET) allows to discharge the pixel anodes after each
read-out cycle.

The 4 mm $\times$ 2.5 mm large detector chips are mounted onto a ceramic
hybrid board.
Each detector pixel is connected to one channel of the CAMEX64B~\cite{camex}
read-out chip.
It provides signal amplification, base line subtraction, and multiplexed 
output.
The digital steering chip TIMEX generates the necessary control signals.
Signal integration times between 2~$\mu$s and the full TTF pulse length 
(800~$\mu$s) can be programmed.

\section{Calibration and Noise Determination}

An absolute energy calibration of each detector pixel is obtained
using mono-energetic photons emitted from a ${}^{55}$Fe source at
5.90~keV (Mn K${}_\alpha$-line, 89.5\%) and 
6.49~keV (Mn K${}_\beta$-line, 10.5\%). 
The X-ray photons enter the detector through the back entrance window 
on the side opposite to the anodes.
Photons at an energy of 6~keV have an attenuation length of 
30~$\mu$m in silicon and are therefore absorbed close to 
the surface of the detector chip.
On average each photon at this energy produces about 1600 electron-hole pairs.
The electrons drift to the anodes, where the charge is collected.
During the drift time of 7~ns the lateral extent of the electron 
cloud increases to about 8~$\mu$m due to diffusion.

\begin{figure}[t]
 \begin{center}
 \mbox{\epsfig{figure=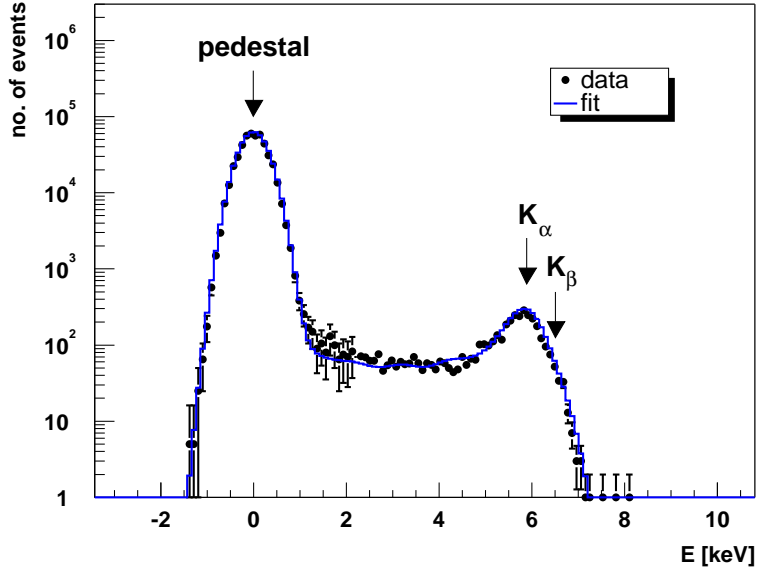,%
               width=0.75\textwidth}}
 \end{center}
 \caption{Measured energy spectrum for pixel 8
          (60~$\mu$m $\times$ 250~$\mu$m).}
 \label{fig_Fe}
\end{figure}

The detector is operated at room temperature and read out with 
a rate of 5~kHz.
For the given activity of the source ($10^6$~Bq) and the integration
time (15~$\mu$s) a photon is registered by the detector 
in 10\% of the read-outs.
The measured energy spectrum for one of the detector pixels is shown
in Fig.~\ref{fig_Fe}.
It can be separated into three parts:
The pedestal peak, 
which dominates the distribution, the signal peak, 
which consists of the K${}_\alpha$ and K${}_\beta$ components, 
and the region in between, 
which is caused by charge sharing between adjacent pixels.
The energy scale, the noise and the diffusion width are determined
with a simultaneous fit to the whole spectrum of Fig.~\ref{fig_Fe} based 
on a model describing the two-dimensional pixel structure.

The location of the pedestal defines the zero-signal level for each pixel.
The pedestal subtraction has already been applied to the data shown in 
Fig.~\ref{fig_Fe}.
The difference between the signal and pedestal peak gives an energy
calibration for each pixel.
The resulting calibration constants differ by at most 10\% for neighbouring 
left and right pixels.
For a 30 $\mu$m photon spot this corresponds to an error in position
reconstruction of at most 0.8 $\mu$m if the signal is not corrected 
according to the different calibration constants.

The Gaussian width of the pedestal peak is mainly caused by one source 
of noise, the leakage current.
Using the calibration one calculates energy resolutions between 222~eV
and 391~eV,
which can be translated into an equivalent noise charge (ENC) between
60 and 106 electrons.
The variation of the noise values are caused by the different pixel sizes
which lead to variable leakage current and capacitance.
Due to the dominant role of the leakage current the energy resolution
strongly depends on integration time and temperature.
Our measurements show that the noise grows proportionally to the 
square root of the integration time and decreases by a factor of two when 
cooling the detector by 16~K.

The number of events with charge division between pixels compared to 
the number of events with the photon signal fully recorded by one pixel 
is directly related to the geometry of the individual pixels and the 
diffusion width of the charge cloud.
This is taken into account by the fit and leads to a Gaussian width 
of the charge cloud at the anode plane of about 8~$\mu$m, 
consistent with our estimations from diffusion.

The common mode noise is defined as the variation of the zero-signal 
level common to all pixels.
For each event the median of the 24 pixel signals is taken as an 
estimate for the common mode value.
It has been found that the common mode shows
no systematic drift and varies only within its standard deviation
of 30 electrons.
Electronic cross talk is seen by the pedestal shift of a pixel 
adjacent to a pixel which registered one photon totally.  
The cross talk amounts to 3\% of the full signal at most.
This corresponds to a reduction in spatial resolution by 6\%
which is acceptable for our application.

\section{Measurement of Spatial Accuracy using a Laser Light Spot}
\label{sec_laser}

The spatial accuracy of the pixel detector is measured 
in a laser test stand \cite{sonja} 
by projecting a laser line-focus onto the pixel structure.
The light emitted by a pulsed laser diode ($\lambda = 670$ nm)
is focused using a micro-focus lens and then 
defocused in one plane using a cylindrical lens.
This setup produces a straight line-focus of about 30 $\mu$m full 
width on the detector surface.
The line-shaped light spot is oriented parallel to the separation 
line of the two pixel rows.
Using a stepping device the light spot is moved across the two 
pixel rows with 0.068 $\mu$m per step to determine its left-right 
symmetry line.

\begin{figure}[b]
 \begin{center}
 \mbox{\epsfig{figure=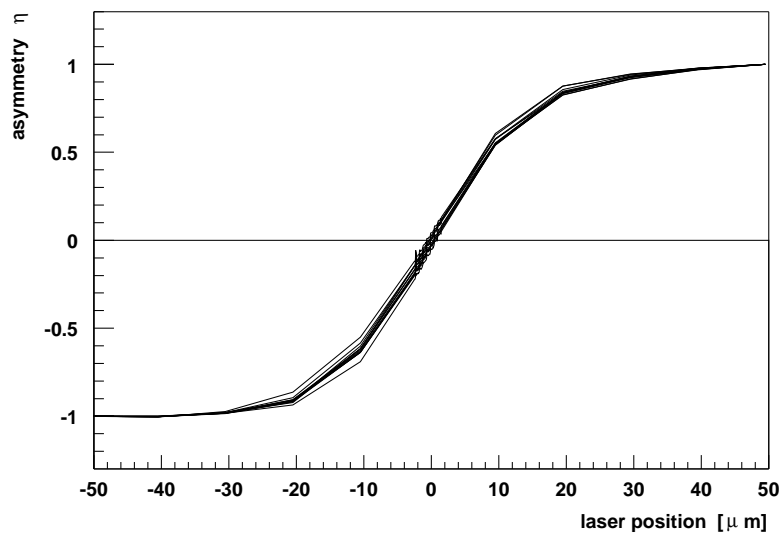,width=0.75\textwidth}}
 \end{center}
 \caption{Asymmetry, $\eta$, versus the position of the laser light spot.
          The results of all 12 pixel pairs are overlayed.}
 \label{fig_laser2}
\end{figure}
In addition to the pedestal correction we subtract a constant 
signal offset proportional to the pixel size to correct for
stray light falling onto the detector.
For each pixel pair the relative difference between the signals
of the right and the left pixel, $\eta = (S_R-S_L)/(S_R+S_L)$, 
is calculated.
The result is shown in Fig.~\ref{fig_laser2} versus the 
position of the laser light spot.
All 12 pixel pairs show the zero crossing of $\eta$ at the same laser 
position within $\pm 1$~$\mu$m.

\begin{figure}[t]
 \begin{center}
 \mbox{\epsfig{figure=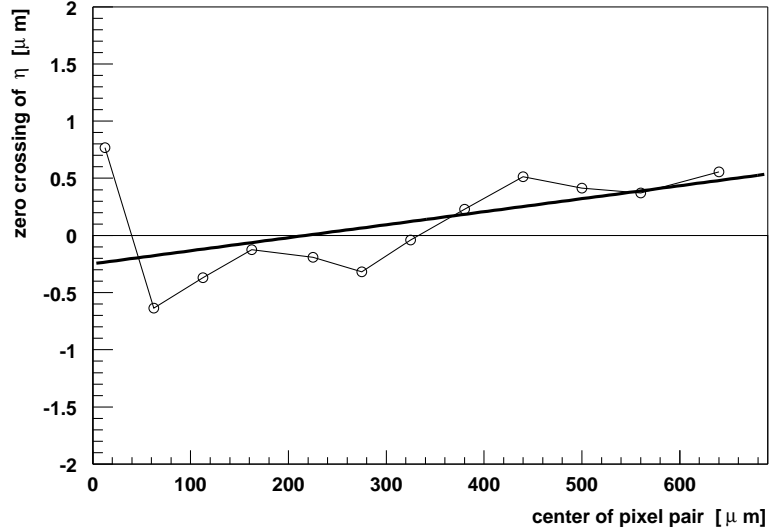,width=0.75\textwidth}}
 \end{center}
 \caption{Zero crossing of $\eta$ versus pixel-pair position.}
 \label{fig_laser3}
\end{figure}
The position of the zero crossing of $\eta$ can be extracted for each 
pixel pair from a straight line fit to the central data points.
In Fig.~\ref{fig_laser3} the resulting zero crossing is plotted versus 
the center position of the corresponding pixel pair.
Obviously, the laser line-focus was tilted by about 1 mrad with respect 
to the center line of the pixel array.
Fitting a straight line to the 12 data points gives us the location of
the laser spot.
The individual measurements scatter with a standard deviation of
0.37~$\mu$m around the reconstructed line.

\begin{figure}[b]
 \begin{center}
 \mbox{\epsfig{figure=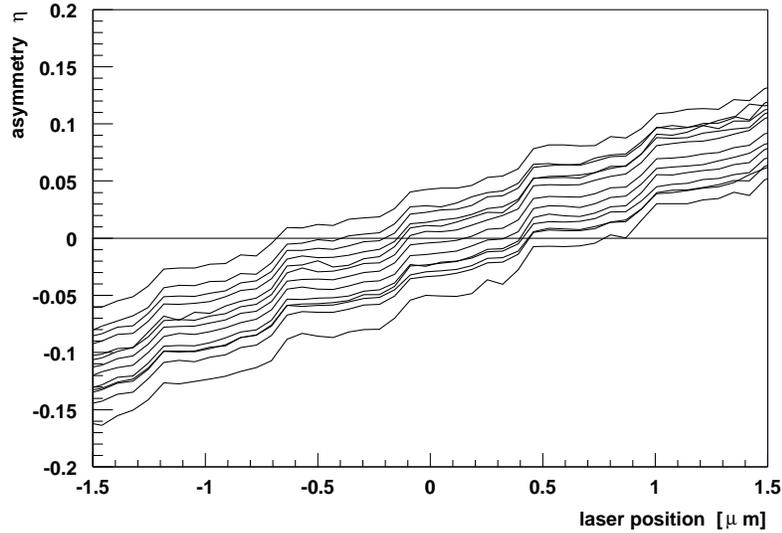,width=0.75\textwidth}}
 \end{center}
 \caption{Closer look at the left-right asymmetry, $\eta$, 
          around zero crossing.}
 \label{fig_laser0}
\end{figure}
For measurements very close to the zero crossing one expects a 
linear dependence of $\eta$ on the laser position.
Fig.~\ref{fig_laser0} shows the behaviour of the measured $\eta$
in this region.
All 12 pixel pairs show the same dependence with the exception of 
different offsets.
We observe a periodic oscillation of 0.5~$\mu$m length which is
caused by the inaccuracy of the stepping device.
As these oscillations are fully correlated one can correct for the
effect and is left with a relative point-to-point resolution of
approximately 0.1~$\mu$m.

\section{Measurement of Spatial Accuracy using a Scanning Electron Microscope}

The detector is installed into the focal plane of a scanning electron 
microscope (SEM).
The SEM produces an electron beam with an energy of 10~keV focused to a 
spot smaller than 1~$\mu$m on the surface of the detector.
The SEM beam current can be adjusted up to 100~$\mu$A at the filament cathode. 
It is significantly reduced by several apertures in the optical system, 
yielding currents below 1~pA at the beam focus.
Secondary emitted electrons from the detector surface are collected and 
amplified by an open multiplier tube.
Its signal is used to display a picture of the detector on a view screen.
The detector hybrid is placed onto a copper plate to remove the heat produced 
by the read-out electronics.
However, the temperature of the detector chips increases by about 15~K while 
the chamber is under vacuum.

The electron beam is scanned across the two pixel rows in parallel to their
separation line.
After each scanning line, the electron beam is displaced by a fixed amount.
The detector read-out is synchronized to the scanning frequency of the SEM,
so that data are taken after each scanned line.
Measurements are made with 618 to 2252 lines per mm.

\begin{figure}[b]
 \begin{center}
 \mbox{\epsfig{figure=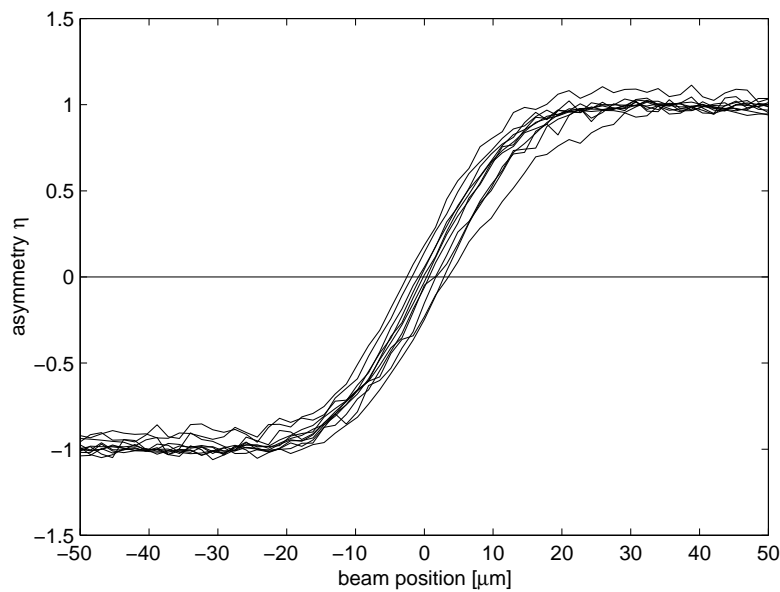,width=0.75\textwidth}}
 \end{center}
 \caption{Asymmetry, $\eta$, versus position of electron beam.
          The results of all 12 pixel pairs are overlayed.}
 \label{fig_sem11}
\end{figure}
\begin{figure}[t]
 \begin{center}
 \mbox{\epsfig{figure=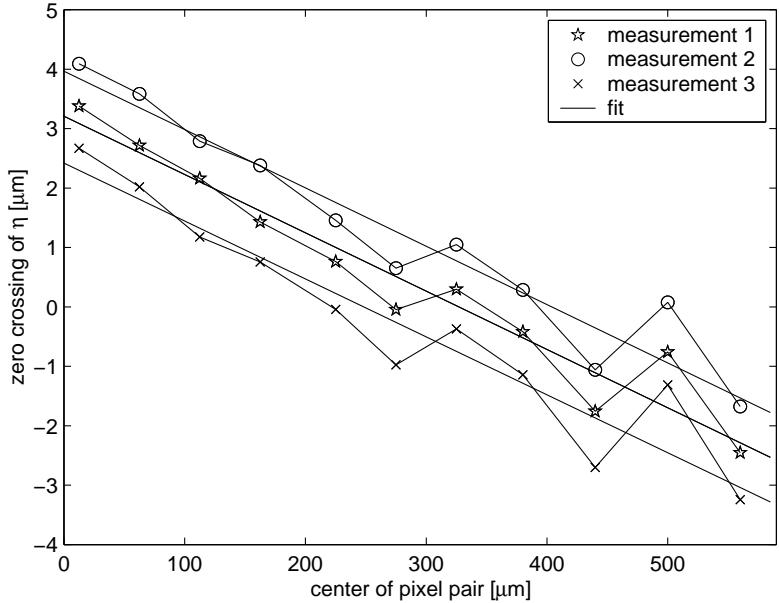,width=0.75\textwidth}}
 \end{center}
 \caption{Zero crossing of $\eta$ versus pixel-pair position}
 \label{fig_sem12}
\end{figure}

Analogous to the previous measurement
the relative asymmetry $\eta$ between right and left pixels is calculated
(see Fig.~\ref{fig_sem11}) and the zero crossings of $\eta$ are extracted.
Fitting a straight line to the central data points gives again the 
zero crossings for the different pixel pairs.
The results of three different scans are shown in Fig.~\ref{fig_sem12}.
The scanning line of the electron beam was tilted with respect to
the symmetry line of the pixel detector by 10 mrad.
The standard deviation of the measured zero crossings from the 
reconstructed scanning line amounts to 0.47 $\mu$m.
The reconstructed scanning lines from three different measurements 
show the same structure and are only shifted with respect to each other.
One concludes that the deviations of the measured zero crossings 
from a  straight line and therefore the limitation in spatial
accuracy of the detector is due to a systematic effect.
The reconstructed symmetry line cannot be directly compared to the
laser measurement of Section~\ref{sec_laser}, because the penetration 
depth of the electrons is much smaller than for laser light.
Additionally, the detector chip was different from the one used 
in the laser test.

\section{Sensitivity to Vacuum Ultraviolet Radiation}

We measured the sensitivity of the detector to vacuum ultraviolet 
(VUV) at the synchrotron radiation facility HASYLAB. 
For this purpose the detector is illuminated with VUV radiation 
in the energy range between 50~eV and 1000~eV which is produced by 
a bending magnet of the electron storage ring DORIS.
The hybrid containing the silicon pixel detector and its read-out
electronics is placed into the vacuum chamber of a reflectometer,
where ultra high vacuum ($10^{-9}$ mbar) had to be established.
The mono-energetic photon beam coming from the monochromator is 
focused onto the center of the pixel detector.
The synchrotron light is pulsed with the 5~MHz bunch frequency 
of DORIS.
The separation of 200~ns between two light pulses 
is much shorter than the integration time of the detector read-out.
Therefore the pixel anodes are discharged at the beginning of 
each integration period which is extended over several synchrotron
light pulses.

\begin{figure}[t]
 \begin{center}
 \mbox{\epsfig{figure=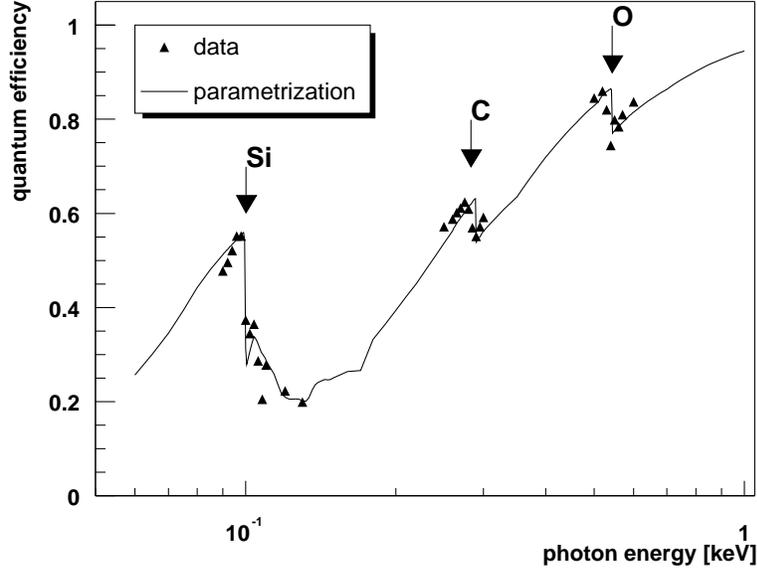,width=0.75\textwidth}}
 \end{center}
 \caption{Quantum efficiency of the detector for VUV radiation.}
 \label{fig_VUV}
\end{figure}

The silicon detector response was corrected to the photo-electron emission 
of one of the focusing mirrors and to a GaAs photo diode as a reference. 
Using the normalized signals, the quantum efficiency can be estimated from 
the measured absorption edges of the relevant elements.
Detailed measurements were done at photon energies in the vicinity of
the absorption edges of silicon (100~eV), oxygen (543~eV), 
and carbon (284~eV). 
For the parameterization of the absorption edges the compilation of 
photon absorption lengths for different elements in Ref.~\cite{henke} is used.
For simplicity we assume that a photon absorbed in the dead layer of the 
detector does not contribute to the signal and all other photons are fully 
registered.
Using the measured heights of the absorption edges,
this model gives the thicknesses of the photon absorbing layers.
These are used to calculate the quantum efficiency in the whole
spectral range from 50~eV to 1000~eV. 
The data points are normalized to this result and both are presented
in Fig.~\ref{fig_VUV}.

\begin{figure}[t]
 \begin{center}
 \mbox{\epsfig{figure=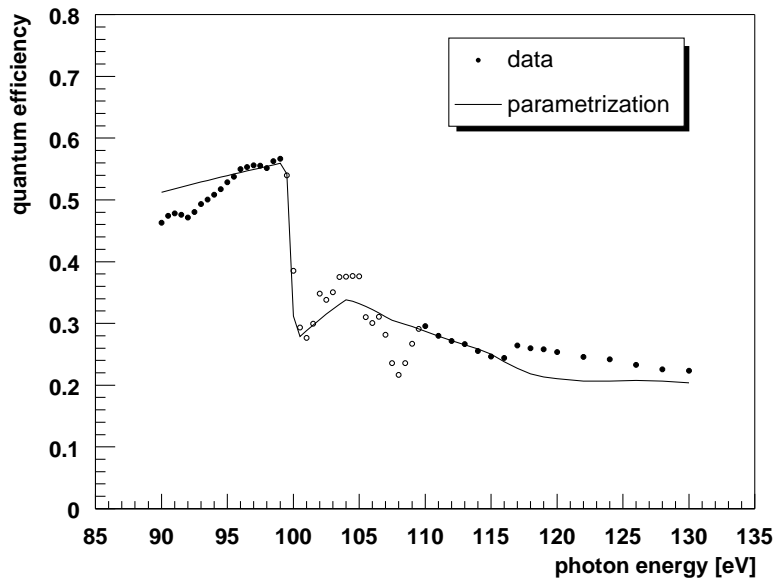,width=0.75\textwidth}}
 \end{center}
 \caption{Quantum efficiency at silicon absorption edge.}
 \label{fig_VUV_det}
\end{figure}

The observed quantum efficiency is explained by the following effects:
The electrical field of the detector diode does not extend up to the
cathode plane, but leaves space for a dead layer with a thickness of the 
order of 30~nm, which has to be penetrated by the photons before they 
enter the sensitive region of the detector.
A 50~nm thin passivation layer of silicon oxide on top 
of the back entrance window leads to further absorption of photons.
From a detailed investigation of the silicon absorption edge 
one can see the effect of the covalent Si-O bond which results in a 
deviation from the absorption edge of pure silicon 
(see Fig.~\ref{fig_VUV_det}).
The origin of an additional carbon contamination which leads to the appearance
of the carbon absorption edge in Fig.~\ref{fig_VUV} is not yet understood.

The quantum efficiency of the detector is larger than 20\%
for photons in the energy range above 100 eV which will be used by the BTM.
Absolute measurements with a similar type of detector have been done 
using a reference diode with known quantum efficiency \cite{hll_ketek}.
The measured quantum efficiencies are comparable with our results, 
but did not show the problem of the carbon absorption edge.

\section{Study of Radiation Hardness}

In the BTM the silicon detector will be operated at a distance of
only 5~mm from the electron beam of the TTF linac.
It can suffer from radiation damages caused by a beam halo
or by scraping a misaligned beam.
One expects that the radiation damages of the silicon 
detectors are dominated by surface effects.

Placing the silicon detector inside an electron microscope 
not only allows to determine its spatial accuracy, but also
gives the opportunity to study radiation hardness against
surface damages.
The detector side containing the pixel anodes and the 
amplifier JFETs should be more sensitive to surface damages
than the back entrance window.
Therefore we place the detector in such way that
the electrons enter the detector opposite to the entrance window.
One of the two pixel rows, including its JFETs, is irradiated 
using the 10~kV electron beam with beam currents of the order 
of several tens of pA.
Irradiation takes place with all operating voltages on, including
the bias voltage.

Before irradiation the detector had been calibrated using the 59.5~keV 
photons of an ${}^{241}$Am source.
During the irradiation procedure,
the beam scan is extended to all pixels and the detector is read out. 
This is done to determine the number of incident beam electrons 
from the measured signal of the undamaged pixels.
From the mean energy loss of a 10~keV electron in silicon oxide
(4.13 keV/$\mu$m) one can then calculate the radiation dose.
Typical dose rates between 0.1~Gy/min and 10 Gy/min have been 
achieved with this method.
The measurement of the beam current was cross checked using a 
faraday cup which delivered consistent results within 20\%.

\begin{figure}[t]
 \begin{center}
 \mbox{\epsfig{figure=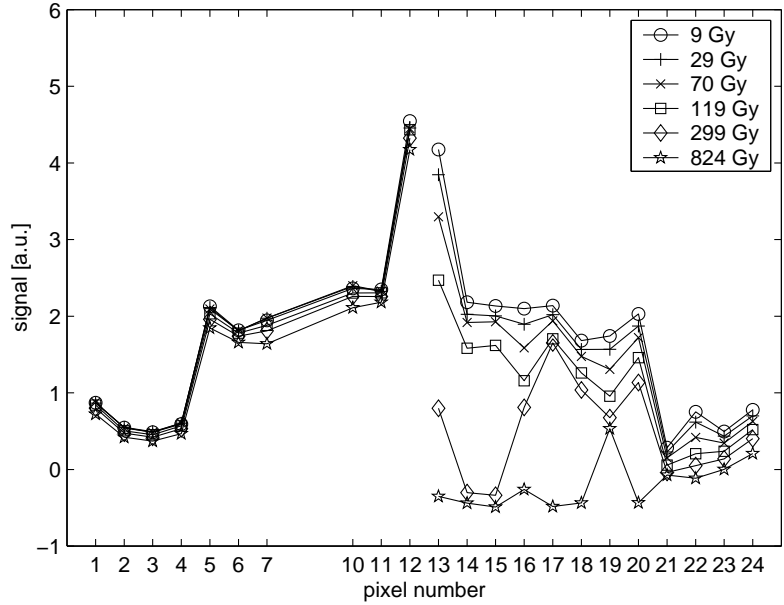,width=0.75\textwidth}}
 \end{center}
 \caption{Signal height of the pixels as function of the radiation dose; 
          pixels 13--24 are irradiated.}
 \label{fig_sem21}
\end{figure}

Several data sets are taken in between the irradiation steps
to determine the change of signal height and noise level.
For this purpose a LED is installed inside the electron microscope
shining onto the back entrance window of the silicon detector.
Fig.~\ref{fig_sem21} gives the measured signal height versus the
pixel number, starting from the small left pixels (1--4), going 
to the large pixels (12,13) and ending with the small right 
pixels (21--24).
The irradiated pixels on the right side show a decreasing 
signal height with increasing radiation dose.
Above radiation doses of about 300~Gy some of the irradiated 
pixels cease to deliver any signal at all.

\begin{figure}[t]
 \begin{center}
 \mbox{\epsfig{figure=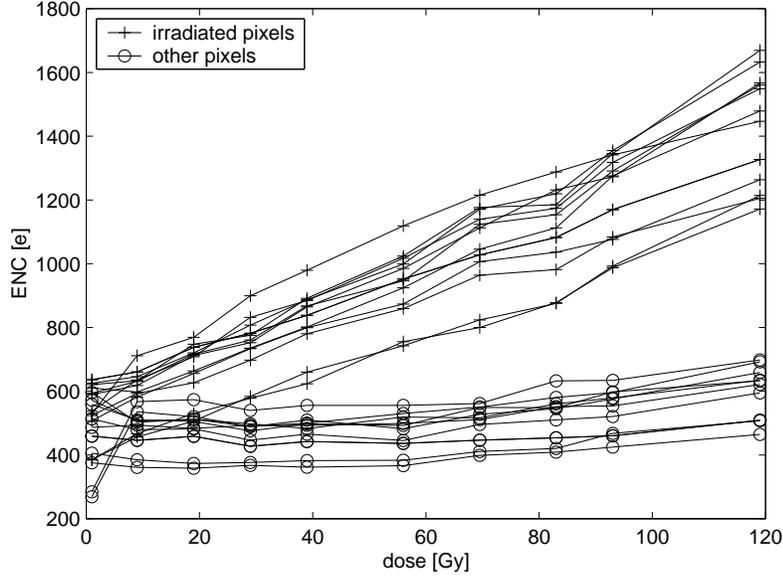,width=0.75\textwidth}}
 \end{center}
 \caption{Eqivalent noise charge as function of the total radiation dose}
 \label{fig_sem22}
\end{figure}

In Fig.~\ref{fig_sem22} the dependence of noise on the radiation
dose is shown. 
One can clearly distinguish between the non-irradiated pixels and 
the irradiated ones.
Whereas the first stayed at the same noise level
the equivalent noise charge of the latter increased by a factor 
of three after a total radiation dose of 120~Gy.

Surface damages include both the creation of oxide charges in the
passivation layer and the generation of inter-band energy levels
at the interface between the silicon bulk and the oxide layer,
the so-called interface states.
The latter inject additional charges and therefore contribute
to leakage current and noise.
The oxide charges lead to a charge up of the SiO${}_2$ layer and
therefore influence the operating voltages of the integrated JFETs.
The signal loss with increasing radiation dose might be caused by
a change of the operation point of the amplifying JFET which results
in a lower gain.
In addition to the amplifying JFET each pixel is connected to a
reset JFET which allows discharging the pixel anode between
detector read-outs.
By recording the dependence of the amplified signal on the applied
gate potential, we measured the gate potential which was necessary 
to close the reset FET.
This voltage had to be increased from $-4$~V to $-6$~V after a
irradiation of 120~Gy.

Further irradiations up to 0.8~kGy were performed, which caused 
the loss of signal in all irradiated pixels.
Recovery of the pixels took place within one week (165 h).
Then the total radiation dose was increased to 4~kGy.
This time the detector could not recover within the following week.
After an in-situ heating of the detector and the read-out electronics 
to 130~${}^\circ$C for 30 minutes all pixels worked again.
The effect can be explained by the removal of the oxide charges due
to the heating process.
Because the interface states cannot be removed by heating the noise
stayed at a high level, a factor of two above the noise of the 
non-irradiated pixels.

\section{Conclusions}

Noise, spatial accuracy and quantum efficiency
of the silicon pixel detector which will be used in the
beam trajectory monitor at TTF-FEL were investigated.
The measured noise values are in the specified range and are
dominated by leakage current.
The systematics of the position measurement was studied 
using a laser line-focus and a scanning electron microscope.
The spatial accuracy is of the order of 0.4~$\mu$m, 
well below the required 1~$\mu$m for the operation
within the beam trajectory monitor.
The sensitivity to vacuum ultraviolet radiation has been 
measured in a synchrotron beam line.
From the observed absorption edges a quantum efficiency 
above 20~\% is estimated at photon energies used for the BTM.
The detector can cope with radiation doses up to 100 Gy.
At the position of the BTM in TTF a radiation dose of the 
order of 1 Gy per week is expected.

\begin{ack}
We are grateful to D.~Vogt for giving us the opportunity to
operate our detector inside an electron microscope.
For his help during the measurements at HASYLAB we would like
to thank M.-A.~Schr\"oder. 
We thank C.~Coldewey, E.~Fretwurst and M.~Kuhnke for fruitful
discussions about radiation hardness of silicon detectors.
\end{ack}

\end{document}